\documentclass[aps,prl,reprint,preprintnumbers,amsmath,amssymb,floatfix]{revtex4-1}

\usepackage{longtable}
\usepackage{morefloats}
\usepackage[dvips]{graphicx}
\usepackage{color}
\usepackage{epsfig,graphicx,amsfonts,amsbsy}
\usepackage{amsmath,amsfonts,amsthm,amssymb}
\usepackage{appendix}
\usepackage{bbm}
\usepackage{makeidx}
\usepackage{url}
\usepackage{verbatim}
\usepackage[bookmarksnumbered,pdfpagelabels=true,plainpages=false,colorlinks=true,linkcolor=blue,citecolor=blue,urlcolor=blue]{hyperref}
\usepackage[rightcaption]{sidecap}
\usepackage{array}
\usepackage{booktabs}
\usepackage{multirow}
\usepackage{bbm}
\usepackage{tabularx}
\usepackage{cancel,soul,ulem}

\newcommand{\bs}[1]{\boldsymbol{#1}}
\newcommand{\vv}[1]{\boldsymbol{#1}}

\graphicspath{{figures/}}

\makeatletter

\begin{document}

\def \cedenna {Centro  de Nanociencia y Nanotecnología CEDENNA, Avda. Ecuador 3493, Santiago, Chile}
\def \fcfm {Departamento de F\'isica, FCFM, Universidad de Chile, Santiago, Chile.}
\def \usach {Departamento de F\'isica, Universidad de Santiago de Chile.}

\title{ Magnonic Hall Effect and Magnonic Holography of Hopfions}
\author{Carlos Saji${}^{1,2}$}
\author{Roberto E. Troncoso${}^{3}$}
\author{Vagson L. Carvalho-Santos${}^{4}$}
\author{Dora Altbir${}^{2,5}$}
\author{Alvaro S. Nunez${}^{1,2}$}

\affiliation{${}^{1}$\fcfm}
\affiliation{${}^{2}$\cedenna}
\affiliation{${}^{3}$School of Engineering and Sciences, Universidad Adolfo Ib\'a\~nez, Santiago, Chile}
\affiliation{${}^{4}$Departamento de F\'isica, Universidade Federal de Vi\c cosa, 36570-900, Vi\c cosa, Brazil}
\affiliation{${}^{5}$\usach}


\date{\today}

\begin{abstract}
Hopfions are localized and topologically non-trivial magnetic configurations that have received considerable attention in recent years. Through a micromagnetic approach, we analyze the scattering of spin waves by magnetic hopfions. We show that the spin waves experience an emergent electromagnetic field  related to the topological properties of the hopfion. We find that spin waves propagating along the hopfion symmetry axis are deflected by the magnetic texture, which acts as a convergent or divergent lens, depending on the spin wave propagation direction. The effect differs for spin waves propagating along the plane perpendicular to the symmetry axis. In the last case, they respond with a skew scattering and a closely related Aharonov-Bohm effect. This allows probing the existence of a magnetic hopfion by magnonic holography.    
\end{abstract}

\maketitle

\noindent{\it Introduction}.- The emergence of particle-like states on different systems \cite{Rajaraman1987, Mermin-RMP} is at the crossroads of several fields in modern physics. From early classical mechanical models\cite{Remoissenet1999}  to contemporary studies of elementary particles \cite{Brown2010}, the nature of such states remains a fertile ground where theorists and experimentalists converge.
Such endeavors rely heavily on the notion that some particle states are protected by their topology \cite{Skyrme1962}. Regarding magnetic systems, in addition to the fundamental interest that topological protection offers to several particle-like systems, the potential for using magnetic quasiparticles in data processing and storage devices \cite{Parkin2008} draws attention also from the applied point of view. Therefore, analyzing several magnetic solitonic states' static and dynamic properties, such as domain walls \cite{Dey2021, Wang2022-1, Landeros2010, Ulloa2016},  vortices, skyrmions \cite{Muhlbauer2009, Fert2017, Schott2017, Huang2022, Du2022, Wang2022, Chakrabartty2022}. Bloch points \cite{Im2019, Tejo2021, Zambrano-Rabanal2022, Li2020, Beg2019, Rana2023}, is one of the main topics of current research.

\begin{figure}[ht]
\includegraphics[width=\linewidth]{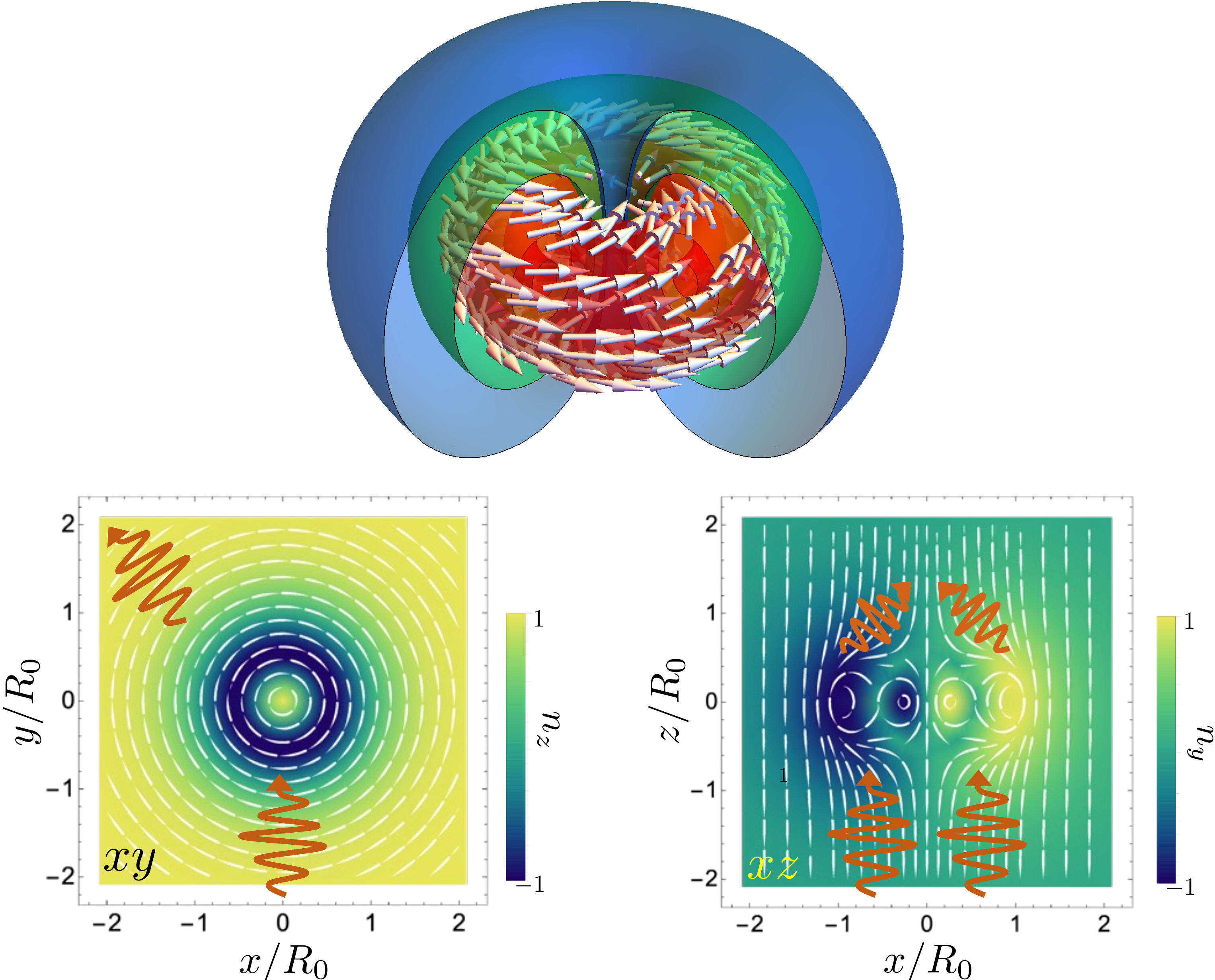}
\caption{(a) Onion-like depiction of a Bloch hopfion where each surface stands for regions with constant $n_z$ (the $z$-axis coincides with the symmetry axis of the hopfion.) The arrows at $n_z=0$ (red surface) represent the magnetization field, $\vv{n}$, as it winds around in space. The outer surfaces correspond to $n_z=0.5$ (green surface) and $n_z=0.9$ (blue surface). $n_z<0$ surfaces are located inside the red surface and are not presented here. The streamlines of the hopfion field $\vv{n}$ in the planes $xy$ and $xz$ are depicted at (b) and (c), respectively. The color code represents the projection of $\vv{n}$ along the normal to each plane. The red wavy lines represent the SWs-hopfion interaction. SWs propagating along the $xy$ plane suffer skew-scattering, while SWs moving along the symmetry axis of the Hopfion are deflected similarly to a converging or diverging lens, depending on the direction of propagation.}
\label{fig: Hopfion cartoon}
\end{figure}

The possibility of engineering nanoparticles with well-controlled shapes, sizes, and magnetic properties allows the nucleation, stabilization, and control of three-dimensional (3D) magnetic textures \cite{3D1,3D2,3D3,3D4}. Amongst the plethora of 3D quasiparticles, we can highlight the magnetic hopfion, which consists of a topological soliton configuration where the magnetization field swirls in a knotted pattern creating a stable structure, see Fig. \ref{fig: Hopfion cartoon}. Despite the basic properties of magnetic hopfions being theoretically studied for more than 20 years \cite{Faddeev1997}, the analysis of their nucleation and stabilization processes \cite{Sutcliffe2018, Liu2022, Rybakov2022, Castillo2021, Corona2023}, and their experimental observation has been reported just recently \cite{Kent2021}. Due to their exciting properties \cite{Wang2019, Gobel2021, Zhang2023, Shen2023}, and localized nature, the control of magnetic hopfions could foster a new era of spintronics devices, dramatically increasing their density and speed while reducing their power consumption.

In parallel to studying the properties of 3D magnetic quasiparticles, the analysis of the interaction between spin waves (SWs) and magnetic textures is, nowadays, a well-established subject of research \cite{Yu2021}. Among the essential phenomena displayed by such interaction, perhaps the most baffling one is the ability of SWs 
to generate a change in the momentum of a magnetic texture \cite{Lan2022}, which can induce its motion along the nanoparticle that holds it. Another interesting behavior related to the interaction between SWs and topological magnetic objects is the emergent magnetic fields generated by skyrmions that can induce magnon Hall effects \cite{Hoogdalem2013}. Additionally, magnonic bands in skyrmion crystals display a topological structure in momentum space akin to those found on the integer quantum Hall effect \cite{RoldnMolina2016}. Regarding the scattering of spin waves in 3D systems, it was shown that the effective field generated by Bloch points on SWs resembles the magnetic field of the exotic Dirac monopole \cite{Elias2014} in such a way that a Bloch point induces a non-trivial structure on the behavior of the SW phases \cite{Carvalho2015}.

This letter presents an analysis of SWs propagating across a magnetic hopfion configuration. Knowledge of such a system includes bound and extended states that inherit much of the topological nature of the underlying texture. It provides  several ways to detect and manipulate magnetic hopfions unambiguously, creating a bridge between the buoyant magnetic field of magnonics \cite{Pirro2021, Yuan20221, Babak20221, Wang2022-2, Roldan2017} and the pursuit of magnetic hopfion creation, detection, and control.

\noindent {\it Structure of magnetic hopfions}.- We consider a chiral magnetic system modeled using the micromagnetic energy functional $\mathcal{E}=\mathcal{E}_{\text{bulk}}+\mathcal{E}_{\text{PMA}}$, where
$
\mathcal{E}_{\text{bulk}}=\int_V \left(J (\mathfrak{D}_{\mu} \vv{n})^{2}/2 -\vv{B}\cdot \vv{n} \right)\ d^{3}\vv{r}.
$
Here, $\vv{n}$ is the normalized magnetization, $\vv{B}$ is the external magnetic field along the $z$ axis, and $\mathfrak{D}_{\mu}=\partial_{\mu} + \kappa\ \vv{e}_{\mu} \times \;$ is the helical derivative, with, $\vv{e}_{\mu}$ being a basis of the spatial coordinates, and $\kappa= D/2J$ is the characteristic helical number, with $J$ the exchange coupling \footnote{Here, following the ideas of Jin \textit{et al.} \cite{Jin_2022} that analyze the magnon-driven dynamics of skyrmions, we assume that the next-nearest neighbors corrections to the exchange energy, that in the continuum limit is ${\cal E}_{\text{nnn}}^{\text{ex}}=C^{\mu \nu}\int \partial_{\mu}^{2}\vv{n}\cdot \partial_{\nu}^{2}\vv{n}d^{3}{\bs r}$, with $C^{\mu \nu}$ corresponding to the respective coupling constant, is negligible in the spin wave Hamiltonian interacting with the hopfion. Therefore, the Heisenberg exchange dominates the interaction, that is, 
${\cal E}_{\text{nnn}}^{\text{ex}} \ll \mathcal{E}_{\text{bulk}}$.} and $D$ the strength of the bulk Dzyaloshinskii-Moriya interaction (DMI) characteristic for noncentrosymmetric materials. The second contribution to the energy functional consists of a perpendicular magnetic anisotropy (PMA), given by $\mathcal{E}_{\text{PMA}}=K_{S}\int_{S}\left(1-\vv{n}_{z}^{2}\right) d^2{\bs r}$ \cite{Hopf1}, where the integral runs over the external surface of the magnetic system and $K_S$ represents the PMA strength. It has been shown that hopfions could be stabilized in confined chiral magnetic systems with perpendicular magnetic anisotropy (PMA) \cite{Hopf1}, geometrical constraints \cite{Castillo2021, Corona2023}, or frustrated exchange interactions \cite{Rybakov2022}. In the former case, anisotropy leads to a magnetization pinning at the upper and bottom layers, preventing the formation of 3D skyrmion tubes \cite{Gobel2021}. The formal description of a hopfion \cite{Hopf1} can be given in terms of the $\mathfrak{su}(2)$ algebra, which works out as a representation of the rotation group $SO(3)$. That is, any element $\vv{\mathcal{R}}=\mathcal{R}_{0}\mathbb{I}_{2}+i\sum_{\mu}\mathcal{R}^{\mu}\hat{\sigma}_{\mu}$, where $\hat{\sigma}_{\mu}$ stands for the $\mu$-th Pauli spin matrix ($\mu=1,2,3$). It can be noticed that these elements satisfy the relation $\mathcal{R}_{0}^{2}+ \sum_{\mu}\mathcal{R}^{\mu}\mathcal{R}_{\mu}=1$ and generate a rotation operator 
with the group action $\hat{\vv{z}}\to\hat{\vv{n}}= \vv{\mathcal{R}}^{-1} \hat{\sigma}_{z} \vv{\mathcal{R}} $, where $\hat{\vv{n}}$ stands for $\sum_{\mu}n^{\mu}\hat{\sigma}_{\mu}$. The field of toroidal hopfions, $\vv{n}_{H}$, are constructed from a rotational symmetry texture $\vv{\mathcal{R}}_{H}(\vv{r})=e^{i f(r){\hat{\vv{r}}}/{r}}$,
where the function $f(r)$ is smooth, monotonic, and satisfies $f(0)=0$ and $f(\infty)=\pi$ 
\footnote{3D magnetic solitons classified under the third homotopy group are characterized by the Hopf index $\mathcal{Q}_{H}$ \cite{KOSEVICH, Zarzuela2019} of the texture that quantifies the linking structure of the magnetization. The Hopf index is formally defined as
$
\mathcal{Q}_{H}=\frac{1}{8\pi^{2}}\int_V \varepsilon_{i j k} \varepsilon_{\alpha \beta \gamma \delta} \mathcal{R}_\alpha \frac{\partial \mathcal{R}_\beta}{\partial x_i} \frac{\partial \mathcal{R}_\gamma}{\partial x_j} \frac{\partial \mathcal{R}_\delta}{\partial x_k} \ d{\bs x},
$
where Einstein convention is assumed for repeated indices. In  particular, for the toroidal hopfion represented by $\vv{\mathcal{R}}_{H}$, we obtain  $\mathcal{Q}_{H}=1$.}.  

Here, we consider a general hopfion characterized by its toroidal and poloidal cycles $p,q$, defined as the 2D-winding numbers on the $(y,z)$ and $(x,z)$ transversal planes, respectively. The profile of the $(p,q)$-hopfion, in spherical  coordinates $(r,\theta,\phi)$, reads
\begin{align}\label{eq: nvec}
\begin{pmatrix}
n_{x}\pm i n_{y}\\ n_{z}
\end{pmatrix}
&=\begin{pmatrix}
\sin(2\chi) e^{\pm i (q\phi-p\Theta + \gamma)}\\
\cos(2\chi)
\end{pmatrix}
\end{align}
where $ \Theta(r,\theta)= \tan^{-1}\left [ \tan(f(r))\cos(\theta) \right ]$ and $\chi(r,\theta)=\sin^{-1}\left [ \sin(f(r)) \sin(\theta)\right ]$.  The parameter $\gamma$ accounts for the helicity ($\gamma=0$ for  N\'eel type hopfions and $\gamma=\pi/2$ for Bloch type hopfions). An illustration of the hopfion texture and their magnetization profiles along $xy$ and $xz$-planes parameterized by Eq. \eqref{eq: nvec} for $\gamma=\pi/2$ are shown in Fig. \ref{fig: Hopfion cartoon}, where white arrows represent the magnetization field.

We also highlight that associated with a magnetic texture is the emergence of a magnetic field defined by the Berry curvature $\vv{B}^{\mathrm{em}}_{\mu}=\epsilon_{\mu \nu \eta}\vv{n}\cdot(\partial_{\mu} \vv{n}\times\partial_{\nu} \vv{n})$.
In terms of this field, the Hopf index can be written as $\mathcal{Q}_{H}={(4\pi)^{-2}}\int_V  \vv{B}^{\mathrm{em}}\cdot \vv{A}^{\mathrm{em}} d{\bs r}$
where  $\vv{B}^{\mathrm{em}}=\nabla \times \vv{A}^{\mathrm{em}}$. Therefore, a remarkable property of hopfions is their vanishing global gyrovector  $\vv{G}_{H}=\int_{V} \vv{B}^{\mathrm{em}} \ d{\bs r}= \vv{0}$, in contrast with the skyrmion case, indicating the absence of the Hall effect in the hopfion dynamics in the rigid body approximation \cite{Liu_PRL_2020}.  

\noindent{\it Numerical simulations.-} Using the GPU-accelerated MuMax$^3$ package \cite{Vansteenkiste2014}, we implement a simulation of the spin wave-hopfion system. Such code solves the Landau-Lifshitz-Gilbert (LLG) equation \cite{Landau, Gilbert} to emulate the dynamics of the magnetization of a ferromagnetic material. We consider a rectangular grid of size $200\times 200 \times 40$ sites with cell sizes of $a=1\ [\mathrm{nm}]$ and periodic boundary conditions.
In addition, we consider a saturation magnetization, $M_{\mathrm{s}}= 100 [k \mathrm{A}/\mathrm{m}]$, an exchange stiffness, $A_{ex}=0.1 [\mathrm{pJ}/\mathrm{m}]$, and $D= 0.05 [\mathrm{m J}/\mathrm{m}^{2}]$, a Gilbert damping parameter $\alpha=0.01$, a surface anisotropy constant $K_{\text{S}} =0.5 [\mathrm{m J}/\mathrm{m}^{2}]$ and volumetric anisotropy, $K_{\mathrm{bulk}}= 5 [\mathrm{ kJ/m^3}]$, for the material parameters. 

The system starts in a configuration described by Eq. (\ref{eq: nvec}) and relaxes toward the final configuration as a function of the external magnetic field. We find that the stability region of the confined hopfion occurs for the magnetic field in the range of $B_{\mathrm{ext}}<B_{\mathrm{c}}=0.05\mathrm{[T]}$. Beyond this threshold value, one observes a Bloch point pair formation, also known as a toron state \cite{Lan2021}. Based on the above-described, to obtain the dynamical properties of the spin wave scattering on hopfions, we consider a magnetic field $B= 0.025[\mathrm{T}]$.  

After stabilizing the magnetic hopfion in the considered system, we study the behavior of SWs propagating along different directions with respect to the obtained hopfions. We consider that SWs are excited by a variable external monochromatic magnetic field to simulate real-time dynamics. Thus, the SW train is obtained from applying an ac magnetic field $\vv{B}=B_{ac}\cos(2\pi f_{ac}t)$, with frequency $f_{ac}=60 [\mathrm{GHz}]$ and strength $B_{ac}=0.05\ [\mathrm{T}]$, in the direction of the planes lying at the boundaries of the system. 

 \begin{figure}[htp]
 \centering
\includegraphics[width=\linewidth]{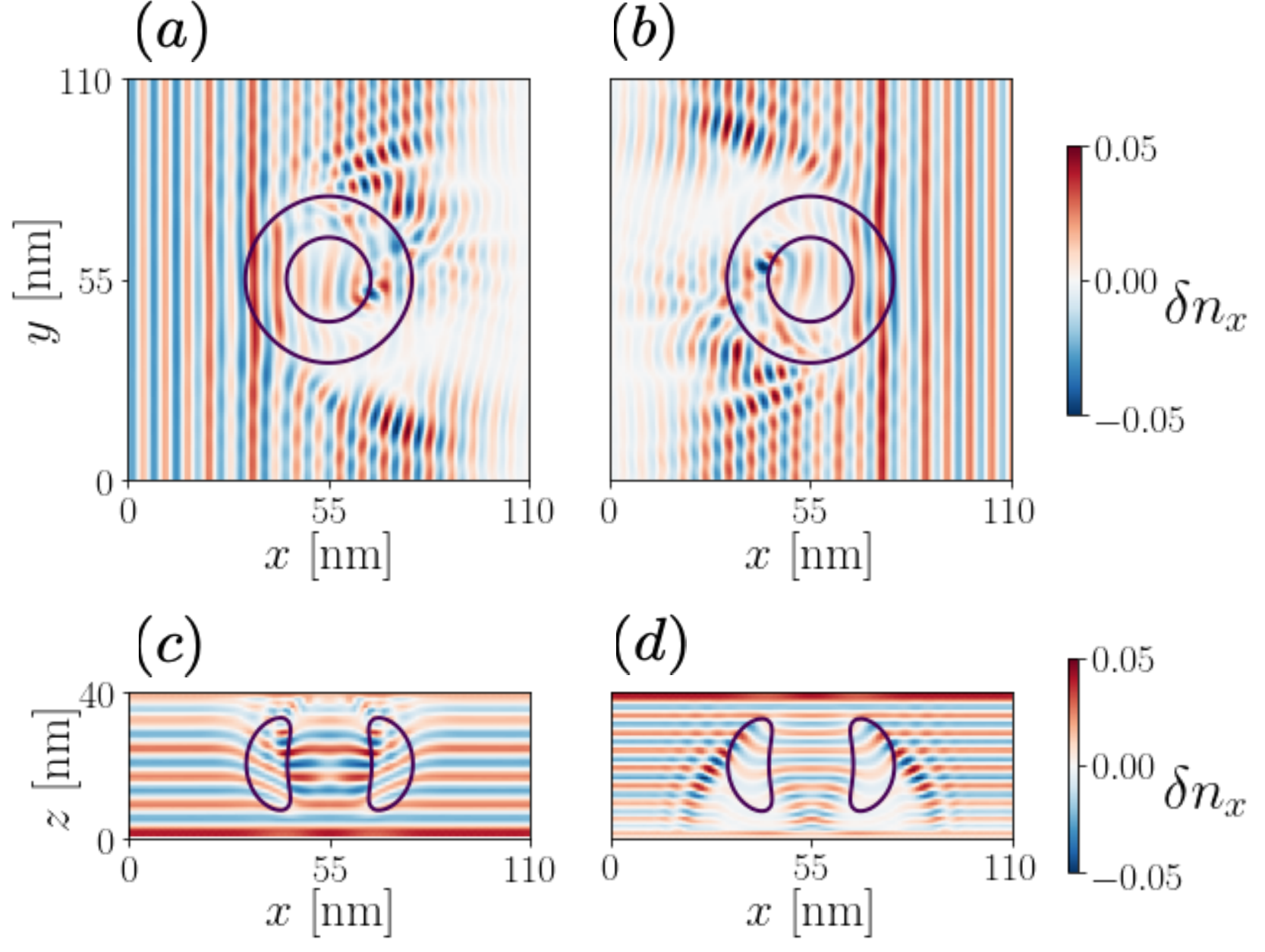}
\caption{Representation of the scattering of a SW by a hopfion obtained by micromagnetic simulations. The top and bottom panel depicts the spatial oscillation $\delta n_{x}(\vv{r},t)=n_{x}(\vv{r},t)-n_{x}(\vv{r},0)$ on the $(X,Z)$ and $(X,Y)$ cross-section of the hopfion, respectively. Here the black contour marked encloses the hopfion isosurface.}
\label{fig: Mumax3 results}
 \end{figure}%

The obtained results for the SW scattering on hopfions are presented in Fig. (\ref{fig: Mumax3 results}), where two main exciting phenomena can be noticed. Suppose the SW propagates along the symmetry axis of the hopfion. In that case, we observe the appearance of an effective-lens behavior, where, depending on the direction of its propagation, SWs converge, as shown in Fig. (\ref{fig: Mumax3 results}.a) or diverge, as depicted in Fig. (\ref{fig: Mumax3 results}.b) after crossing the hopfion. On the other hand, when the SW propagates in the plane perpendicular to the symmetry axis, the hopfion acts as a source of skew scattSupposeing, similar to the Magnus effect, leading to a magnon-Hall effect. Again, the SW's direction after crossing the hopfion depends on the propagation direction, see Figs. (\ref{fig: Mumax3 results}.c and d).

\noindent{\it  Spin waves around magnetic Hopfions}.- To understand the behavior described above, we perform an analytical account of the spin waves that considers them as a small perturbation around the hopfion background field $\hat{\vv{n}}_{H}=\vv{\mathcal{R}}^{-1}(\vv{r})\hat{\vv{z}} \vv{\mathcal{R}}(\vv{r}) $. 
The spin connection associated with this gauge transformation is defined by
$\vv{\mathcal{A}}_{\mu} = \vv{\mathcal{R}}^{-1} \mathfrak{D}_{\mu}\vv{\mathcal{R}}=\vv{\mathcal{R}}^{-1} \partial_{\mu}\vv{\mathcal{R}} + \kappa \vv{e}_{\mu} $ (see Eq. (\ref{eq: gauge field}) in supplementary material). 
Now, we introduce excitations of the hopfion using the Holstein-Primakoff transformation, or equivalently, by performing the linearization  $\hat{\vv{z}}\to \hat{\vv{z}}' \approx \psi \hat{\sigma}_{+} + \psi^{*} \hat{\sigma}_{-}  + (1- 2|\psi|^{2} )\hat{\sigma}_{z}$, with $\hat{\sigma}_{\pm}=\hat{\sigma}_{x}\pm i\hat{\sigma}_{-}$. Hence, expanding the energy functional up to second order in $\Psi=(\psi,\psi^{*})^{t}$, and using the identity $\partial_{\mu} (\vv{\mathcal{R}}^{-1}\hat{\vv{n}}_{0}\vv{\mathcal{R}})=\vv{\mathcal{R}}^{-1}(\partial_{\mu}\hat{\vv{n}}_{0}+[\vv{\mathcal{A}}_{\mu},\hat{\vv{n}}_{0}] )\vv{\mathcal{R}}$, we write the energy of the SWs in the presence of a hopfion as $
\mathcal{E}_{\mathrm{SW}}[\Psi]={J}\int \Psi^{\dagger}(\vv{r})\mathcal{H}_{\mathrm{SW}}\Psi(\vv{r}) \ d^{3}\vv{r}/2 ; ,
$
with the Bogoliubov-de Gennes ($\mathrm{BdG}$) Hamiltonian given by
\begin{align}
&\mathcal{H}_{\mathrm{SW}} =\nonumber \\
&\begin{pmatrix}
(i\partial_{\mu}-\mathcal{A}_{\mu}^{Z}(\vv{r}))^{2}+\mathcal{U}(\vv{r})  &  \mathcal{V}(\vv{r}) \\ 
 \mathcal{V}^{*}(\vv{r})  & (-i\partial_{\mu}-\mathcal{A}_{\mu}^{Z}(\vv{r}))^{2}  +\mathcal{U}^{*}(\vv{r})
\end{pmatrix}\label{eq: Effective Hamiltonian}.
\end{align}
The corresponding potentials are obtained and presented in Eqs. (\ref{eq: potential vector}-\ref{eq: potential anomalous}) of Supplemental Material (SM). The result presented in Eq. (\ref{eq: Effective Hamiltonian}) shows that magnons interacting with a hopfion are exposed to an effective magnetic field, defined by the Berry curvature $\vv{B}^{\mathit{eff}}=\nabla \times \vv{\mathcal{A}}^{Z}$, that affects their dynamics.  In the regime where $\kappa$ can be neglected, it coincides with the emergent magnetic field of the texture $\vv{B}^{\mathrm{em}}$. Under these statements, the magnon spin current is determined as
\begin{equation}
\mathcal{J}_{\mu}= \Psi^{\dagger} (-i\sigma_{z}\partial_{\mu}+\mathcal{A}^{Z}_{\mu})\Psi,
\label{eq: magnon current}
\end{equation}
meaning that magnons are coupled to a hopfion through the $z$-component of the spin connection $\vv{\mathcal{A}}_{\mu}$. 

\noindent {\it Spin-wave scattering}.- We now analyze the effect of the effective magnetic field over the magnon through scattering experiments. Then, the semi-classical approach results in a useful approximation for highly energetic magnons. Let us consider a spin wave  $\displaystyle{\psi(\vv{r},t)=e^{-i\omega t+\vv{k}\cdot \vv{r}} }$ incoming from the $-\vv{k}$ direction. In a semi-classical approach, the motion equation reads $\dot{\vv{\mathsf{r}}}=\vv{\mathsf{p}}/m^{*}$, where $m^{*}=\hbar M_{\rm s}/(2 J a^{2}\gamma_{0})$ determines the magnon mass and a Lorentz force (arising from the effective magnetic field) dominates the momentum evolution 
$
\dot{\vv{\mathsf{p}}}=\vv{B}^{\mathit{eff}}(\vv{\mathsf{r}})\times \dot{\vv{\mathsf{r}}} - \vv{\nabla} \mathcal{U}(\vv{\mathsf{r}}),
$
where $\mathcal{U}(\vv{\mathsf{r}})$ relates with the potential vector (see Eq. (\eqref{eq: potential normal}) in SM).

Another insightful way to describe the magnon scattering on a hopfion is by considering its finite toroidal moment, given by  
$
\vv{\mathsf{t}}=\int d^3 r[\vv{r} \times \vv{B}^{\mathit{eff}}(\vv{r})]/2=\mathsf{t}_{z} \hat{\vv{z}}.
$ One can notice that the hopfion's toroidal moment points along the direction of its symmetry axis, and its magnitude is proportional to the total magnetic charge contained within the hopfion. Therefore, by using the Belavin-Polykov ansatz, $\sin (f(r))={2rR_{0}}/({r^{2}+R_{0}^{2}})$, one obtains  $\mathsf{t}_{z}= {3}\pi^{3} R_{0}^{2}/2$. A direct effect of the toroidal magnetic moment arises by considering a wavefront propagating along the $z$-axis. Magnons propagating with velocity $v_{z} \hat{z}$, are subjected to an average axis-radial force $\vv{F}_{\rho}=2p\,{\sin^{2}(f)f'}\sin(\theta) v_{z}\vv{\rho}/{r}$. Due to its explicit dependence on $v_z$, the effective force exerted by hopfions on magnons can be attractive or repulsive depending on the direction of the SW propagation. Therefore, the scattering problem is reduced to the SW scattering on a system composed of a skyrmion and an anti-skyrmion with the same radius ($\approx R_{0}$). Moreover, assuming that the effective magnetic flux is highly concentrated at the inner region of the torus isosurface, the magnon scattering can be seen as the  intersection of the cyclotron orbit in that region. In this context, following a similar argument as in Ref. \cite{Yu2019}, the magnon deflection Hall angle is given by $\theta_{\mathrm{Hall}} \approx \pi/\rho_{\mathrm{cyc}}$, where $\rho_{cyc}\approx m^{*}v R_{0}/4p$ is the cyclotron radius. 

\begin{figure}[htb]
 \centering
\includegraphics[width=\linewidth]{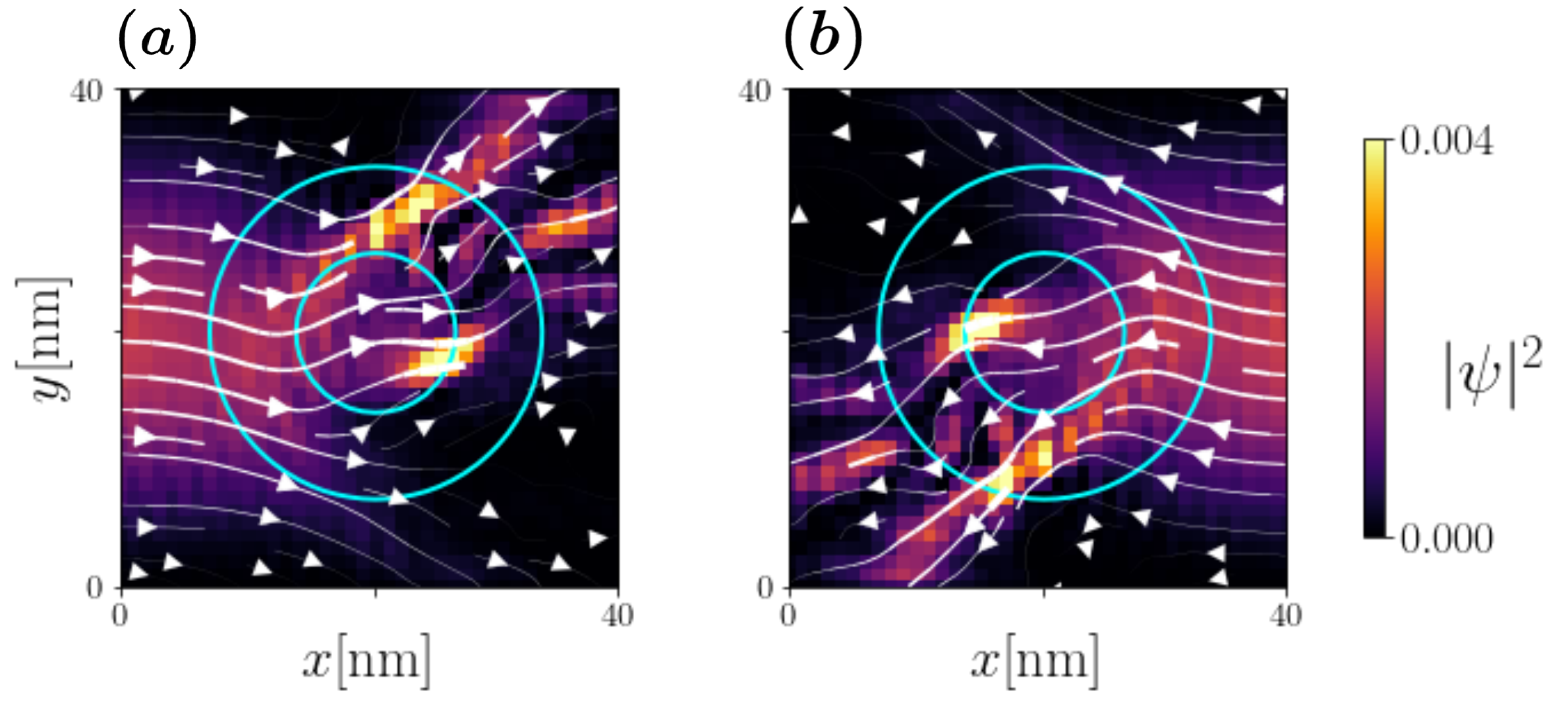}
\includegraphics[width=\linewidth]{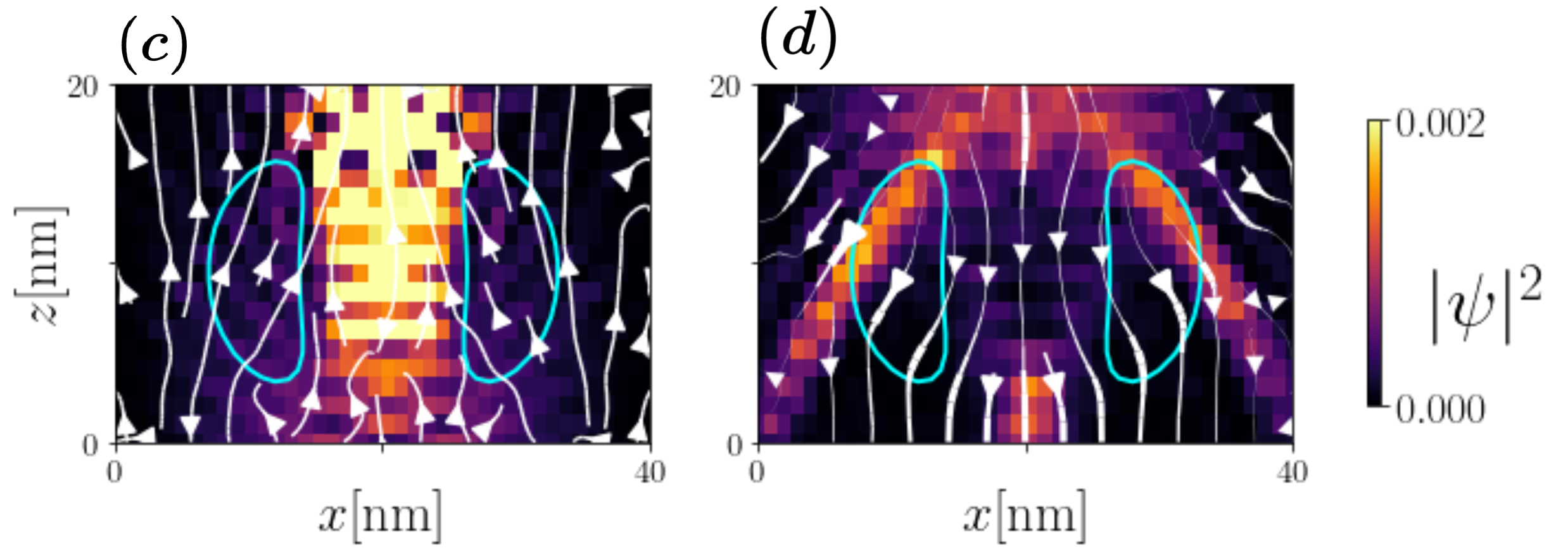}
\caption{Scattering states associated with incoming spin waves from different directions. The lines correspond to streamlines of the magnon current in Eq.(\ref{eq: magnon current}). Color intensity represents the local density  of magnons $\rho=|\psi|^{2}$. Top panel: Spin waves propagating along the plane perpendicular to the hopfion's symmetry axes. The solid line is a reference to the location of the hopfion. Bottom panel: Spin waves propagating along the hopfion's symmetry axis.  The solid line illustrates the location and dimensions of the hopfion.}
\label{fig: Kwant results}
 \end{figure}

To support our conclusions, we solved Eq. (\ref{eq: Effective Hamiltonian}) by using the package kwant \cite{kwant}, which consists of a Python library specialized in quantum transport. While its specific purpose is related to the electronic properties of quantum systems, it is possible to use it in the context of generic wave propagation. We looked for solutions with boundary conditions in the form of incoming plane waves from different directions. The obtained results evidence that when spin waves propagate along the plane perpendicular to the hopfion’s symmetry axes, a skew scattering effect can be appreciated with opposite signs according to the direction of propagation of the spin waves, either from left to right, see Fig. (\ref{fig: Kwant results}.a) or from left, see Fig. (\ref{fig: Kwant results}.b). On another side, when spin waves propagate from bottom to top, the interaction with the hopfion is equivalent to traveling to a divergent lens as seen in Fig. (\ref{fig: Kwant results}.c). Waves propagating from top to bottom  experience a convergent lens shown in Fig. (\ref{fig: Kwant results}.d). These results agree with our theoretical analysis, which predicts that The focal length of such an effective magnonic lens depends on the toroidal moment of the hopfion. Additionally, the magnon Hall effect and the magnonic lens behavior are qualitatively equivalent to the results obtained from micromagnetic simulations.  
 
Finally, an exciting effect that arises from the effective flux piercing each plane at the  cross-section of the system is the magnon Ahronov-Bohm effect. It acquires a simple form in the $xy$ plane where the phase difference between two interfering arbitrary magnon paths enclosing to the hopfion can be calculated considering the integral of the potential vector $\Phi_{\mathit{flux}}=\oint \vv{A}\cdot d\vv{r}$  over a circle of radius much larger than the system size. From the adopted theoretical model, we find for the net flux $\Phi_{\mathit{flux}}= 2\kappa R_{0}$ corresponding to twice the flux determined for the skyrmion case \cite{Han}. This flux opens up the possibility of magnon interference experiments \cite{Carvalho2015} that will attest to the presence of the hopfion and readily lead to the development of magnon holographic techniques.

\noindent{\it Discussion.-} This letter presented the analysis of SWs propagating across a hopfion configuration through micromagnetic simulations and an analytical model based on small perturbations, described as a linear expansion around the hopfion ground state. The natural way the hopfion affects the propagation of linear excitations is through Berry's phases that depend on the geometrical details of the texture. These effects are cast in terms of an effective magnetic field that acts on the spin waves. Following this trail, we arrived at two main conclusions corroborated by the micromagnetic and analytical calculations. First, in the regime of short wavelength, there is an effective-lens behavior for propagation along the symmetry axis of the hopfion. The focal length of such an effective magnonic lens depends simply on the toroidal moment of the hopfion. In the plane perpendicular to the axis of symmetry, the effect of hopfion is to act as a source of skew scattering, leading to a magnon-Hall effect. Second, we have shown that the propagation of waves of any wavelength will be affected by an Aharonov-Bohm effect, extremely sensitive to the relative plane in which the propagation will take place. We argue that as a possible set-up for holographically detecting the hopfion and its features. It might serve as a platform to implement magnonic holographic devices suitable for data processing \cite{Khitun2013}.

\noindent{\it Acknowledgments.-} Funding is acknowledged from Fondecyt Regular 1190324, 1220215, and Financiamiento Basal  para  Centros  Cient\'ificos  y  Tecnol\'ogicos  de  Excelencia AFB220001.  C.S. thanks the financial support provided by ANID National Doctoral Scholarship. V.L.C.-S. Thanks to the Brazilian agencies CNPq (Grant No. 305256/
2022-0) and Fapemig (Grant No. APQ-00648-22) for financial support. V.L.C-S also acknowledges Universidad de Santiago de Chile and CEDENNA for hospitality.

\bibliography{bibliography}

\onecolumngrid
\section{Supplemental Material}

In this Supplemental Material, we show the details of the calculations of the effective connection field acting on the SWs and the solutions to the short-wavelength equations.

\subsection{Emergent magnonic gauge fields}

According to the $\mathbb{C}\mathbb{P}^{1}$ representation, the hopfion vector field can be written as  $n_{\mu}=\langle \vv{z}  |\hat{\sigma}_{\mu}|\vv{z} \rangle$ where the spinor $| \vv{z} \rangle=(z_{1},z_{2})^{T}$  satisfies $|z_{1}|^{2}+|z_{2}|^{2}=1$ . We define the $(p,q)$-degree hopfions, in spherical coordinates $(r,\theta, \phi)$, with the following spinor:\begin{align}
	|\vv{z} \rangle =\left (\cos(\chi)\  e^{i p \Theta} ,  i\sin(\chi)\ e^{i q\phi} \right )^{T} \label{eq: pqhopfion}
\end{align}
where the function $\chi(r,\theta),\Theta(r,\theta)$ are defined in the main text.
In order to compute the geometrical spin connection, we use the relation  
\begin{align}
	\mathcal{A}_{\mu}^{k}\hat{\sigma}_{k}&= i \hat{\mathcal{R}}^{-1}\partial_{\mu}\hat{\mathcal{R}} , \quad \hat{\mathcal{R}} =\begin{pmatrix}
		z_{1} & z_{2}\\ 
		-z_{2}^{*} & z_{1}^{*}
	\end{pmatrix}\,.
	\label{eq: gauge field}
\end{align}
Thus substituting Eq. \ref{eq: pqhopfion} in Eq. \ref{eq: gauge field}, we obtain
\begin{align*}
	\mathcal{A}^{Z}_{\mu} &= q \sin^{2}(\chi) \partial_{\mu}\phi - p \cos^{2}(\chi) \partial_{\mu}\Theta  \\
	\mathcal{A}^{X}_{\mu}+i\mathcal{A}^{Y}_{\mu} &= e^{i\Xi} \left ( -\partial_{\mu}\chi - \frac{i}{2}\sin(2\chi)\partial_{\mu}\Xi\right )
\end{align*}
where $\Xi=q\phi-p\Theta$. Now, adopting a spherical coordinate basis and adding the Dzyaloshinkii-Moriya contribution ($\partial_{\mu}\to\mathfrak{D}_{\mu}=\partial_{\mu}+\kappa \vv{e}_{\mu} \times$) , the full expressions for the effective potential and magnetic fields yield 

\begin{align}
	\vv{\mathcal{A}}^{Z}&=p f'(r)\cos(\theta)\ \vv{r} - p \frac{\sin(2f(r))}{2r}\sin(\theta)\ \vv{\theta} + q\frac{\sin^{2}(f(r))}{r}\sin(\theta) \ \vv{\phi}+\kappa \vv{n}_{H} \\
	\vv{B}^{\mathit{eff}} =\nabla \times \vv{\mathcal{A}}^{Z}&= q \frac{2\sin^{2}(f)}{r^{2}}\cos(\theta)\ \vv{r} - q\frac{\sin(2 f)f'}{r}\sin(\theta)\ \vv{\theta} + p \frac{ 2\sin^{2}(f)  f'}{r}\sin(\theta)\ \vv{\phi} + \kappa \nabla \times \vv{n}_{H}
	\label{eq: potential vector}
\end{align}

\begin{figure}[htb]
	\centering
	\includegraphics[width=5cm]{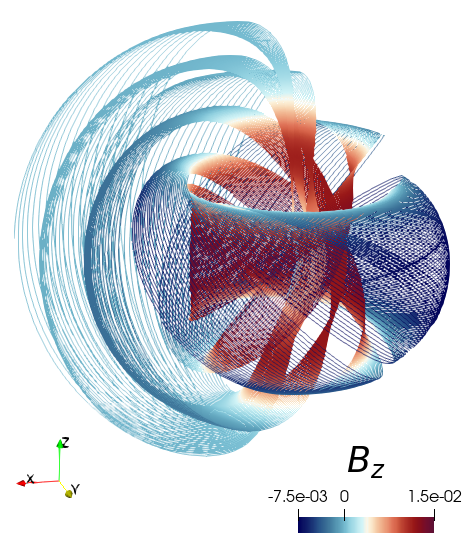}
	\caption{Knotted streamlines of the effective magnetic fields $\vv{B}^{\mathit{eff}}$  for a hopfion with $p=1,q=1$ . The linking number between each pair of lines corresponds to the Hopf index $\mathcal{Q}_{H}=1$. }
	\label{fig: figure2}
\end{figure}

Finally, we calculate the magnon potential by following the treatment given in \cite{Han}
\begin{align}
	\mathcal{U} &= - \frac{1}{2} \sum_{\mu}\left ( |\mathcal{A}^{X}_{\mu}|^{2}+ |\mathcal{A}^{Y}_{\mu}|^{2}  \right )  + \vv{B}\cdot \vv{n}_{H} \label{eq: potential normal}\\
	\mathcal{V} &= \frac{1}{2} \sum_{\mu}\left ( \mathcal{A}^{X}_{\mu} +i \mathcal{A}^{Y}_{\mu}  \right )^{2}   \label{eq: potential anomalous}
\end{align} 
The rotational symmetry of the spin connection can be analyzed as follows. Let $\mathcal{O}\in SO_{3}$ be a rotation in such a way  that the rotated hopfion profile is $\hat{\vv{n}}_{H}'(\vv{r})=\hat{\vv{n}}_{H}(\mathcal{O}\vv{r})=\vv{\mathcal{R}}'^{-1}\hat{\vv{k}}\vv{\mathcal{R}}'$, where $\vv{\mathcal{R}}'=\vv{\mathcal{O}}^{-1}\vv{\mathcal{R}}\vv{\mathcal{O}}$. Therefore, the potential transforms according to $\vv{\mathcal{A}}_{\mu}'=\vv{\mathcal{O}}^{-1}\vv{\mathcal{A}}_{\mu}\vv{\mathcal{O}}$.  On the other hand, an axis inversion of the hopfion ($\vv{z}\to -\vv{z}$) changes the sign of the effective field ($\vv{B}_{\mathit{eff}}\to -\vv{B}_{\mathit{eff}}$ ).

\subsection{Magnon Tight Binding Model}

The Hamiltonian operator in chiral magnet systems is given by 
\begin{align*}
	H=- \sum_{\vv{r},,\vv{r}'}   & \frac{S}{2}\left( J\vv{n}_{\vv{r}} \cdot \vv{n}_{\vv{r}'} +   \vv{D}_{\vv{r},\vv{r}'} \cdot (\vv{n}_{\vv{r}} \times \vv{n}_{\vv{r}'})\right) + \sum_{\vv{r}}\vv{B}\cdot \vv{n}_{\vv{r}}
\end{align*}
Let $\vv{n}$ be the background state magnetization and consider the orthonormal basis defined by the rotated basis $\vv{n}=\mathcal{R}\vv{z},\vv{e}_{\pm}=\mathcal{R}(\vv{x}\pm i\vv{y})$. Thereby, using the HP transformation, it provides the spin waves Hamiltonian in the form:
$$
\mathcal{E}^{SW}=\sum_{\vv{r},\vv{r}'}  \vec{\Psi}_{\vv{r}}^{\dagger} \mathcal{H}_{\vv{r},\vv{r}'}^{\mathrm{SW}}\vec{\Psi}_{\vv{r}'}
$$
where the hopping matrix  elements are given by the Hessian of the energy 
\begin{align*}
	\mathcal{H}_{\vv{r},\vv{r}',a,b}^{\mathrm{SW}} &=\frac{\delta^{2} H }{\delta \vv{n}_{\vv{r}} \delta \vv{n}_{\vv{r}'} }=-J\vv{e}_{\vv{r}}^{a}\cdot \vv{e}_{\vv{r}'}^{b}  - \vv{D}_{\vv{r},\vv{r}'}\cdot (\vv{e}_{\vv{r}}^{a}\times \vv{e}_{\vv{r}'}^{b} ) - \delta_{\vv{r},\vv{r}'} \mathcal{E}_{\vv{r}}
\end{align*}
and with $\mathcal{E}_{\vv{r}}=-\sum_{\vv{r}'}\left( J\vv{n}_{\vv{r}} \cdot \vv{n}_{\vv{r}'} +   \vv{D}_{\vv{r},\vv{r}'} \cdot (\vv{n}_{\vv{r}} \times \vv{n}_{\vv{r}'})\right) - \vv{B}\cdot \vv{n}_{\vv{r}}$ is the energy density of the background texture. 

The above Hamiltonian is used to calculate the magnon scattering by a static hopfion in the numerical experiments with kwant \cite{kwant}.

\subsection{High energy approximation}

In the same line as the electron scattering by a magnetic field\cite{Sergey2021}, Born approximation provides a direct method to calculate scattering amplitude for magnons in collision with a hopfion as follows
\begin{align*}
	\mathcal{S}(\vv{p}',\vv{p}) &=\left \langle \vv{p}' \left | \vv{A}^{Z}(\vv{r})\cdot\hat{\vv{p} }+\hat{\vv{p} } \cdot\vv{A}^{Z}(\vv{r}) + \mathcal{U}(\vv{r}) \right | \vv{p}\right \rangle \\
	&=\widetilde{\vv{A}^{Z}}(\vv{p}-\vv{p}')\cdot(\vv{p}+\vv{p}') + \widetilde{\mathcal{U}}(\vv{p}-\vv{p}')\,, 
\end{align*}where 
the tilde denotes the Fourier transforms. In particular, the $0$-order terms are calculated directly as $\widetilde{\vv{A}^{Z}}(0)= \int \vv{A}^{Z}(\vv{r}) d^{3}\vv{r}=\int \vv{r}\times\vv{B}^{\mathit{eff}} d^{3}\vv{r} =\vv{\mathcal{T}}$, which consists of the toroidal magnetic moment of the field $\vv{B}^{\mathit{eff}} $. Additionally, $\widetilde{\mathcal{U}}(0)=\int \mathcal{U}(\vv{r})\ d^{3}\vv{r}=-\mathcal{E}_{0}$, where $\mathcal{E}_{0}$ is the energy of the hopfion in the ground state. Hence, at first order, we obtain a non-reciprocal scattering of spin waves propagating along the $\hat{\vv{z}}$-axis as follows
$$
\mathcal{S}(\vv{p}',\vv{p}) \approx \mathcal{T}_{z} (p_{z}+p_{z}') - \mathcal{E}_{0}\,.
$$

The angle deflection of the scattered magnons can be calculated using the eikonal approximation. Since the limit of high-energy magnons dominates the derivatives $\partial_{\mu}\psi$ in the Lagrangian, we can neglect the scalar potentials contribution. Thereby, the total phase shift $\delta_{\infty}$ accumulated along the trajectory $x=b,y=0,z=vt$ is evaluated as: 
\begin{align*}
	\delta_{\infty}(\vv{b}) &\approx \int_{-\infty}^{\infty} \vv{\mathcal{A}}(\vv{b}+\vv{x})\cdot d\vv{z} \\
	&= (-b \sin(\theta_{b}) +\kappa b^{2}\sin(2\theta_{b}) ) F(b)
\end{align*}
where $F(b)=\int_{-\infty}^{\infty} \frac{(f(\sqrt{x^{2}+b^{2})})^{2}}{x^{2}+b^{2}} \ dx$ . The deflection angle $ \alpha(b)$ turns out to be:
$$
\alpha(b)= \frac{2}{k} \frac{\partial \delta_{\infty}(b)}{\partial b}\approx \frac{2}{k} (- \sin(\theta_{b}) +2\kappa b\sin(2\theta_{b}) ) F_{0}(b)\,.
$$
Finally, the deflection angle for a direct collision is  $\alpha=  \alpha(b=0)$ 
$$
\alpha= -\frac{2}{k} \sin(\theta_{b}) F_{0}(0) \approx -\frac{4\pi \sin(\theta_{b}) }{kR}\,.
$$

\end{document}